\documentclass[twocolumn]{aastex61}
\usepackage{comment}

\newcommand{\source}{\mbox{IRAS 18566$+$0408}}
\newcommand{\lsun}{\mbox{L}_{\odot}}
\newcommand{\msun}{\mbox{M}_{\odot}}

\newcommand{\trece}{^{13}\mbox{CO}}

\newcommand{\dcodu}{^{12}\mbox{CO(2--1)}}
\newcommand{\tcodu}{^{13}\mbox{CO(2--1)}}

\newcommand{\sosccc}{\mbox{SO(6$_{5}$--5$_{4}$)}}
\newcommand{\vlsr}{\mbox{v}_{\mbox{\tiny lsr}}}

\newcommand{\chcn}{\mbox{CH}_{3}\mbox{CN(12}_{k}\mbox{--11}_{k}\mbox{)}}
\newcommand{\com}{\mbox{compact }}
\newcommand{\vex}{\mbox{very-extended }}
\newcommand{\kms}{\mbox{km s$^{-1}$}}

\newcommand{\grad}{^{\mbox{\tiny o}}}

\shorttitle{SMA observations toward IRAS 18566}
\shortauthors{Silva et al.}
\begin{document}
\title{SMA observations of the hot molecular core \source}

\correspondingauthor{Andrea Silva}
\email{andrea.silva@tufts.edu}

\author{Andrea Silva}
\affil{Harvard-Smithsonian Center for Astrophysics, 60 Garden Street, Cambridge, MA 02138, USA}
\affil{Department of Physics and Astronomy, Tufts University, Medford, MA 02155, USA}
\affil{National Astronomical Observatory of Japan, National Institutes of Natural Sciences, 2-21-1 Osawa, Mitaka, Tokyo 181-8588, Japan}

\author{Qizhou Zhang}
\affil{Harvard-Smithsonian Center for Astrophysics, 60 Garden Street, Cambridge, MA 02138, USA}

\author{Patricio Sanhueza}
\affil{National Astronomical Observatory of Japan, National Institutes of Natural Sciences, 2-21-1 Osawa, Mitaka, Tokyo 181-8588, Japan}

\author{Xing Lu}
\affil{National Astronomical Observatory of Japan, National Institutes of Natural Sciences, 2-21-1 Osawa, Mitaka, Tokyo 181-8588, Japan}

\author{Maria T. Beltran}
\affil{INAF - Osservatorio Astrofisico di Arcetri, Largo E. Fermi 5, I-50125, Italy}

\author{Cassandra Fallscheer}
\affil{Central Washington University, 400 E University Way, Ellensburg, WA 98926, USA}

\author{Henrik Beuther}
\affil{Max-Planck Institute for Astronomy, K\"onigstuhl 17, 69117 Heidelberg, Germany}

\author{T.K. Sridharan}
\affil{Harvard-Smithsonian Center for Astrophysics, 60 Garden Street, Cambridge, MA 02138, USA}

\author{Riccardo Cesaroni}
\affil{INAF - Osservatorio Astrofisico di Arcetri, Largo E. Fermi 5, I-50125, Italy}

\begin{abstract}

We present Sub\-mi\-lli\-me\-ter Array (SMA) observations toward the high-mass star-forming region $\source$. Observations at 1.3 mm continuum and in several molecular line transitions were performed  in the com\-pact (2\farcs4 angular resolution) and very-extended ($\sim$0\farcs4 angular resolution) configurations. 
The continuum emission from the compact configuration shows a dust core
 of 150  $\msun$, while the very-extended configuration reveals a dense (2.6 $\times$ 10$^7$ cm$^{-3}$) and compact 
 ($\sim$4,000 AU) condensation of 8 $\msun$. 
We detect 31 molecular transitions from 14 species including CO isotopologues, SO, CH$_{3}$OH, OCS, and CH$_{3}$CN. Using the different $k$-ladders of the CH$_{3}$CN line, we derive a rotational temperature at the location of the continuum peak of 240 K. 
The $\dcodu$, $\tcodu$, and $\sosccc$ lines reveal a molecular outflow at PA$\sim$135\arcdeg\ centered at the continuum peak. 
The extended $\dcodu$ emission has been recovered with the IRAM 30 m telescope observations. 
Using the combined data set, we derive an  outflow mass of 16.8 $\msun$. 
The chemically rich spectrum and the high rotational temperature confirm  that \source\ is harboring a hot molecular core.  We find no clear velocity gradient that could suggest the presence of a rotational disk-like structure, even at the high resolution observations obtained with the very-extended configuration.

\end{abstract}


%=====================================================================
\section{Introduction}

The study of the formation of high-mass stars, especially in the early evolutionary phases, has been observationally challenging due to the large distances involved (d$\geq$1 kpc), complex cluster environments (n$_{\star}\gtrsim$100 pc$^{-3}$ ), and short evolutionary timescales (t$_{\mbox{\tiny KH}} \leq$10$^{4}$ yr for a O-type star) in comparison with the study of low-mass stars.  Early stages  of massive star formation have numerous observational signatures, including molecular masers \citep[e.g.,][]{wang2006, cyganowski2012}, outflows \citep[e.g.,][]{zhang2001, zhang2005, beuther2002b, qiu2008}, chemically evolved regions \citep[e.g.,][]{wang2014, sanhueza2012, sanchez17}, hot molecular cores \citep[e.g.,][]{Garay1999,Beltran16}, hyper/ultracompact H II regions \citep[e.g.,][]{kurtz2000}, and ionized winds and jets \citep[e.g.,][]{guzman2014,Guzman16}

Hot molecular cores (HMCs) contain gas and dust near or around sites of recent star formation \citep{cesaroni2005a}.
 They are characterized by small sizes ($\leq$0.1 pc), high gas densities (n$_{\mbox{\tiny H$_{2}$}}$$\geq$10$^{7}$ cm$^{-3}$), and high-temperatures ($T_{k}\geq100$ K) \citep{Garay1999,kurtz2000}. Unlike H II regions, these objects present weak or undetectable free-free emission. The lack of free-free emission has been interpreted as due to an intense mass accretion phase that quenches the development of an H II region \citep{walmsley1995, keto2002, keto2003}. Therefore, HMCs are likely the precursors of H II regions.
The hot core phase is thought to last about 10$^{5}$ yr \citep{vandishoeck1998} to 10$^{6}$ yr \citep{garrod2006} and represents the most chemically-rich phase of the interestellar medium, characterized by the presence of complex organic molecules such as CH$_{3}$OH, CH$_{3}$CN, HCOOCH$_{3}$, CH$_{3}$OCH$_{3}$, and CH$_{3}$CH$_{2}$CN \citep[e.g.,][]{beuther2005, beuther2006}.
The observation of HMCs is important for understanding the evolutionary sequence and physical conditions of massive star formation. Due to the compact and dense nature of HMCs, high angular resolution observations using molecular lines with high critical densities and excitation temperatures are crucial to uncover their physical conditions and to search for accretion disks. In recent years, the search for accretion disks in HMCs has been very fruitful, as summarized in the review by 
\cite{Beltran16}, and Keplerian disk candidates have been found \citep[e.g.,][]{Johnston15,Chen16,Ilee16,Beuther17}

$\source$ is a high-mass star-forming region located at a kinematic distance of 6.7 kpc \citep{sridharan2002} with a total far-infrared luminosity of $\sim$8$\times$10$^{4}$ $\lsun$ \citep{zhang2007}  that comes from a single compact ($<$5\arcsec) dust continuum source, indicating the presence of an embedded O8 ZAMS high-mass star \citep{sridharan2002}.
Weak radio continuum emission was detected at 3.6 cm \citep[flux density of 0.7 mJy; ][]{carral1999} and at 2 cm
 \citep[flux density of 0.7 mJy;][]{araya2005}, suggesting that this source is  in a phase prior to the development of a bright ultracompact H II region. Very recent observations with the VLA in the 6 and 1.3 cm wavelength bands resolve the radio continuum source in four components consistent with an ionized jet \citep{Hofner17}.  
Dense gas traced by CS and CH$_{3}$CN, as well as (sub)mm continuum emission, has been observed \citep{bronfman1996, sridharan2002, beuther2002a, williams2004}. NH$_{3}$(1,1) and (2,2) emission were detected first using single-dish telescopes by \citet{miralles1994}, \citet{molinari1996}, and  \citet{sridharan2002}, and later imaged using the VLA \citep{zhang2007}.

Current star formation in the region is also evident from the presence of 22 GHz H$_{2}$O, 6.7 GHz CH$_{3}$OH, 8 GHz H$_{2}$CO, and 6.03 GHz OH  maser emission \citep{miralles1994, slysh1999, beuther2002a,beuther2002b, beuther2002c, araya2007, almarzouk2012}, as well as the presence of molecular outflows detected in SiO and  NH$_{3}$ \citep{zhang2007} at PA=135\arcdeg.  \cite{araya2007} suggest the presence of an ionized jet, inferred by emission at 4.5 $\mu$m 
from {\it Spitzer} and 3.6 cm continuum from VLA, roughly in the direction of the molecular outflow. This jet has been 
recently confirmed by \cite{Hofner17} at $\sim$0\farcs3 angular resolution. 
 Compact and narrow  NH$_{3}$(3,3) line emission features are found associated with the outflow \citep{zhang2007}, likely arising from weak population inversion similar to maser emission \citep{zhang1995, zhang1999}. 
\citet{araya2007} detected emission at 7 mm of a torus-like structure (PA=44$^{o}$) almost perpendicular to the observed ionized jet (PA$\sim 100^{o}$). 

In order to study the physical and kinematical properties of the high-mass star-forming region IRAS 18566+0408, we carried out sub-arcsecond angular resolution observations with the Submillimeter Array in both
 continuum and molecular line emission. $\source$ has a bipolar molecular outflow, which is expected to be associated to 
 a disk. Our main goal is to search for the presence or absence of disk-like structure(s) around the 
 high-mass star(s) driving the molecular outflow. This paper is organized as follow: in $\S$\ref{sect_observations} we summarize the observations, in $\S$\ref{sect_results} we present the results,  and in $\S$\ref{sect_discuss} we discuss and conclude the main findings of the investigation.

\begin{figure*}[!htbp]
\begin{center} 
\includegraphics[angle=0,scale=0.65]{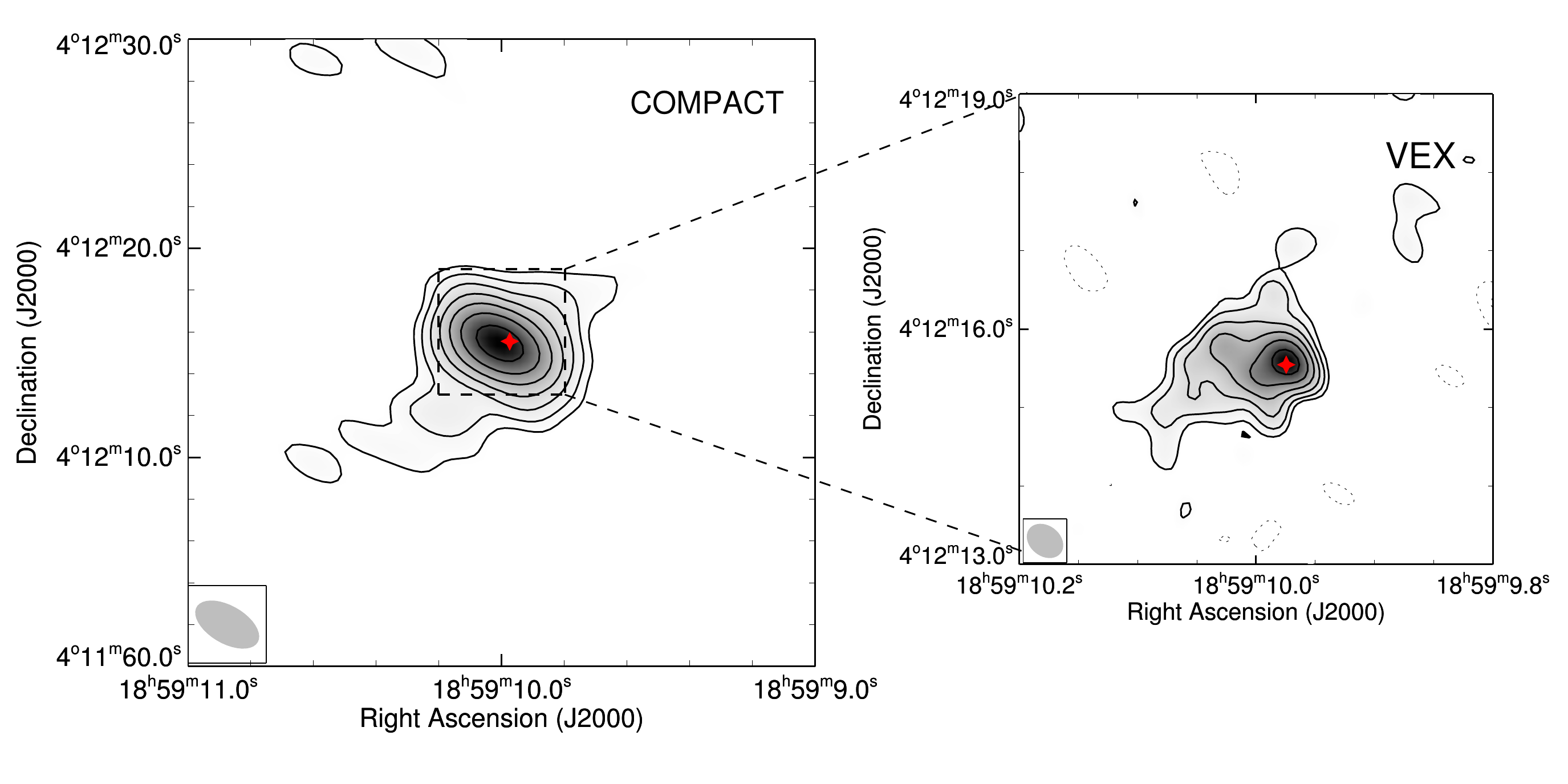}
\caption{ Left: 1.3 mm continuum emission observed with  the \com  configuration ($\theta$=2\farcs4).  The contour levels start at 3$\sigma$ (1$\sigma$=3.71 mJy beam$^{-1}$) and increase in steps of $\log$($\sigma$)=0.2. 
Right: Continuum emission obtained with the \vex configuration ($\theta$=0\farcs4). Same contour levels are plotted  (1$\sigma$=1.1 mJy beam$^{-1}$).  The position of the peak continuum emission detected with the \vex configuration is  indicated with the red star.
The synthesized beams are shown at the bottom left corners.   Negative features are shown with contours in dashed lines.  The peak offset between the two configuration images is less than the smallest beamsize, therefore they are coincident.    \label{fig_continuum}}
\end{center}
\end{figure*}

\vspace{1cm}

\section{Observations}\label{sect_observations}

Observations of $\source$ were carried out with the Sub-millimeter Array\footnote{The Submillimeter Array is a joint project between the Smithsonian
Astrophysical Observatory and the Academia Sinica Institute of Astronomy and
Astrophysics, and is funded by the Smithsonian Institution and the Academia
Sinica.}  (SMA) \citep{ho2004} using two different  configurations. Compact-array observations were taken on July 9, 2007 at 2\farcs4 angular resolution. Very-extended configuration data were obtained on August 3, 2008 at 0\farcs4 angular resolution.
Two spectral sidebands, of 2 GHz wide and separated by 10 GHz, covered the frequency ranges of 219.4--221.3 GHz and 229.3--231.2 GHz.  The receivers were tuned to 230.538 GHz in the upper sideband with a channel spacing of 0.53 $\kms$.
The primary beam of the SMA at 225 GHz is 56\arcsec.
The phase reference center of the observation was set to the position $\alpha$(J2000)=18$^{\mbox{\tiny h}}$59$^{\mbox{\tiny m}}$09$^{\mbox{\tiny s}}$.99 and
$\delta$(J2000)= 04$\grad$12\arcmin15\farcs7.
Quasars J1751+096 and J1830+063 were observed to monitor the time-dependent antenna gains in both configurations.
For the $\com$ configuration the flux was scaled by observing the asteroid Vesta. For the $\vex$ configuration the flux calibration was achieved by observing MWC349
and the Jovian satellite Callisto.
For bandpass calibration,  quasars 3C 273 and 3C 454.3 were observed for the $\com$ and $\vex$ configuration, respectively. 
The data from different configurations were calibrated separately using the IDL-based MIR package. 
The imaging and data analysis were done with the  {\sc miriad} software package \citep{sault1995}.

 The continuum emission was produced by averaging the line-free channels in the visibility
domain. The 1$\sigma$ noise for the continuum emission were 3.7 and 1.1 mJy beam$^{-1}$ for the 
compact and very-extended configurations, respectively. For spectral lines, the 1$\sigma$ noise were 
55 and 45 mJy beam$^{-1}$ per channel for the compact and very-extended configurations, respectively 
(see Table~\ref{tb_lines} for detected lines).

Since the extent of the shortest baseline in the SMA observations is 12 m, corresponding to $\sim$30\arcsec\ at 225 GHz, more extended spatial structures are resolved out. This spatial filtering mainly affects  the low density tracers, such as $\dcodu$ at velocities close to the systemic velocity of the cloud (v$_{\mbox{\tiny lsr}}$ = 85.2 $\kms$).

To recover the missing short spacing information, we observed the $^{12}$CO(2--1) emission at $\sim$11\arcsec\ with the IRAM 30 m telescope located in Pico Veleta, Spain. A 140\arcsec\ $\times$ 140\arcsec region centered at the same position of the interferometric observations was observed. The receiver was tuned to 230.538 GHz. The spectra have 0.4 $\kms$ spectral resolution and were processed using {\sc class}.
The combination of the SMA and the IRAM data was performed using {\sc miriad} \citep{zhang2000}.

\section{Results}\label{sect_results}

\begin{deluxetable*}{lcccccccc}[!htbp]
\tabletypesize{\scriptsize}
\tablecaption{Parameters of the Observed Lines.\label{tb_lines}}
\tablewidth{0pt}
\tablehead{\colhead{Line}&\colhead{Freq.}& \colhead{E$_{\mbox \small{u}}$/k$_{\mbox{\tiny B}}$}&\colhead{Config}&\colhead{$S_{ul}$}&\colhead{1$\sigma$\tablenotemark{a}} \\ \colhead{}& \colhead{[GHz]}& \colhead{[K]}&\colhead{}&\colhead{}&\colhead{$\left[ \frac{\mbox{mJy}}{\mbox{beam}} \frac{\mbox{km}}{\mbox{s}} \right]$} }
\startdata
CH$_{3}$CH$_{2}$CN(24$_{2,22}$-23$_{2,21}$)&219.505&135.6&COM&23.83& 0.21\\
C$^{18}$O(2-1)&219.560&15.8&COM&2.00&  1.19\\
HNCO(10$_{2,8}$-9$_{2,7}$)&219.737&231.1&COM+VEX&9.60 & 0.16 \\
HNCO(10$_{0,10}$-9$_{0,9}$)&219.798&58.0&COM+VEX&9.73& 0.20\\
H$_{2}^{13}$CO(3$_{1,2}$-2$_{1,1}$)&219.909&32.9&COM&2.67& 0.16\\
SO(6$_{5}$-5$_{4}$)&219.949&35.0&COM+VEX&5.95& 0.32\\
CH$_{3}$OH(8$_{0,8}$-7$_{1,6}$)E&220.079&96.6&COM+VEX&1.21&0.19  \\ 
H$_{2}$CCO(11$_{1,11}$-10$_{1,10}$)&220.177&76.4&COM&10.91& 0.15 \\ 
$^{13}$CO(2-1)&220.399  &11.0&COM+VEX&2.00& 0.91\\
CH$_{3}$CN(12$_{8}$-11$_{8}$)&220.476  &525.6&COM+VEX&6.67&0.10 \\
CH$_{3}$CN(12$_{7}$-11$_{7}$)&220.539  &418.6&COM+VEX&7.92& 0.10\\
HNCO(10$_{1,9}$-9$_{1,8}$)&220.585  &101.5&COM+VEX&9.52& 0.18\\
CH$_{3}$CN(12$_{6}$-11$_{6}$)&220.594  &325.9&COM+VEX&9.00&0.13 \\
CH$_{3}$CN(12$_{5}$-11$_{5}$)&220.641  &247.4&COM+VEX&9.92&  0.12\\
CH$_{3}$CH$_{2}$CN(25$_{2,24}$-24$_{2,23}$)&220.661&143.0&COM& 24.80&0.23 \\
CH$_{3}$CN(12$_{4}$-11$_{4}$)&220.679  &183.1&COM+VEX&10.67& 0.28 \\
CH$_{3}$CN(12$_{3}$-11$_{3}$)&220.709  &133.2&COM+VEX&11.25& 0.22\\
CH$_{3}$CN(12$_{2}$-11$_{2}$)&220.730  &97.4&COM+VEX&11.6&  0.20\\
CH$_{3}$CN(12$_{1}$-11$_{1}$)&220.743  &76.0&COM+VEX&11.92& 0.10\\
CH$_{3}$CN(12$_{0}$-11$_{0}$)&220.747  &68.9&COM+VEX&12.00& 0.11 \\
CH$_{3}$OH(15$_{4,11}$-16$_{3,13}$)E&229.589  &374.4&COM+VEX&1.61& 0.19\\
CH$_{3}$OH(8$_{-1,8}$-7$_{0,7}$)E&229.759&89.1&COM+VEX&1.77& 0.17\\
CH$_{3}$OH(19$_{5,15}$-20$_{4,16}$)A$^{+}$&229.864&578.6&COM&2.00&  0.11\\
CH$_{3}$OH(3$_{-2,2}$-4$_{-1,4}$)E&230.027  &39.8&COM+VEX&0.26&  0.19\\
CH$_{3}$OCH$_{3}$(17$_{2,15}$-16$_{3,14}$)&230.234  &147.7&COM&2.98& 0.19\\
O$^{13}$CS(19-18) &230.317 &110.5&COM&19.00& 0.11\\
$^{12}$CO(2-1) &230.538&16.6&COM+VEX&2.00&  2.38\\
OCS(19-18)&231.061  &110.9&COM+VEX&19.00& 0.22\\
$^{13}$CS(5$_{0}$-4$_{0}$)&231.221  &26.7&COM+VEX&10.00&0.25 \\
CH$_{3}$OH(10$_{2,9}$-9$_{3,6}$)A$^{-}$&231.281&165.3&COM+VEX&0.93&0.25 \\
CH$_{3}$CH$_{2}$CN(27$_{0,27}$-26$_{1,26}$)&231.312&157.7&COM+VEX&23.17&0.12 \\
\enddata
\tablecomments{ Molecular line, rest frequency, energy in the upper state, SMA configuration in which the line was detected, line strength, and sigma for the integrated intensity maps. }
\tablenotetext{a}{The noise level per channel is of the order of 50 mJy beam$^{-1}$. Panels in Figure~\ref{fig_spectra_compact_1} and \ref{fig_spectra_compact_2} have different integration ranges in velocity.  
In this table, we summarize the 1$\sigma$ level used in each panel for each molecule. }
\end{deluxetable*}

\begin{figure*}[!htbp]
\begin{center}
\includegraphics[angle=0,scale=0.72]{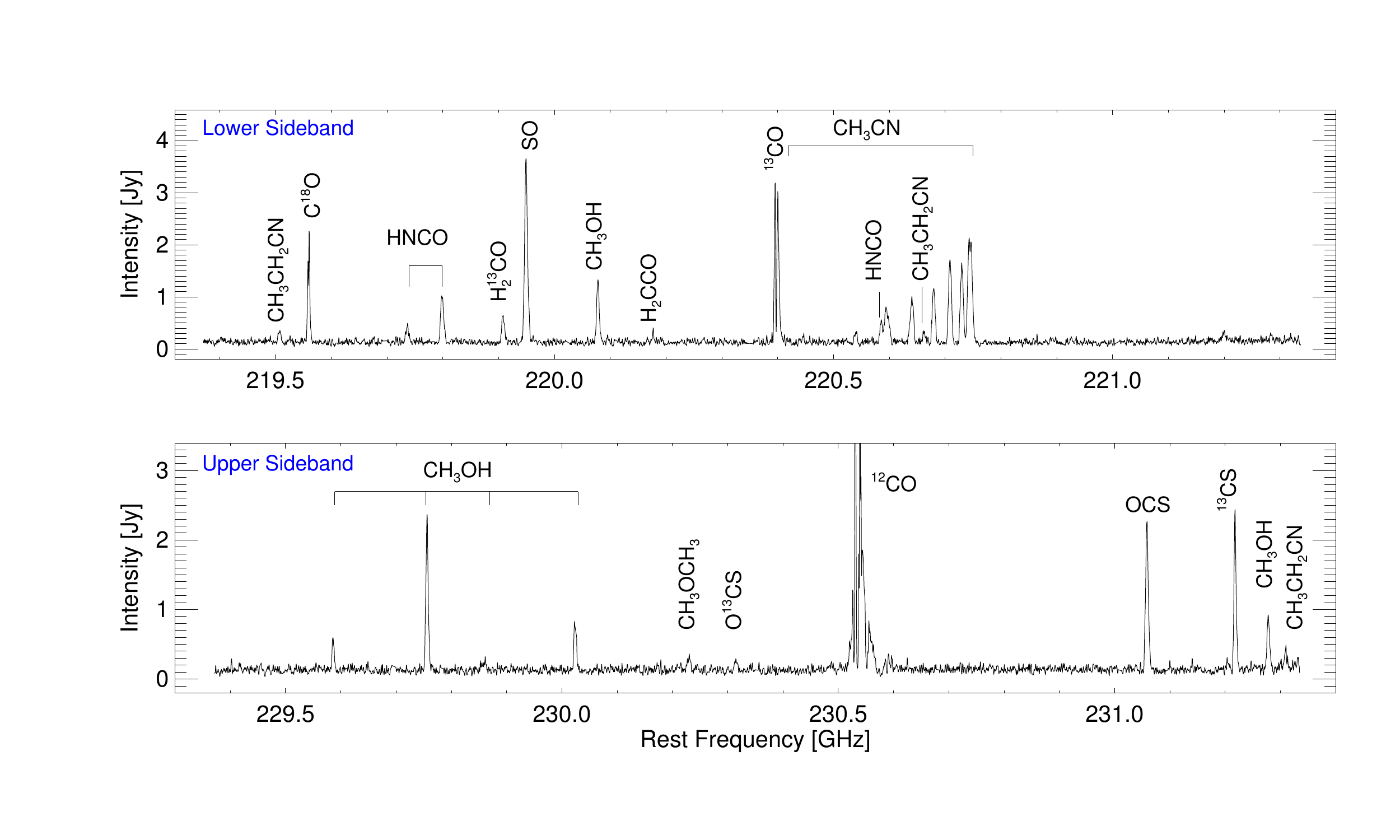}
\caption{Upper and lower sideband spectra time-averaged over all baselines  for the $\com$ configuration.  The spectral resolution is 0.5 km s$^{-1}$ per channel. The identified molecular lines are marked.  The peak emission of $\dcodu$ is 5.2 Jy.  
\label{fig_sbcompact}}
\end{center}
\end{figure*}

\subsection{Continuum Emission}

Figure  \ref{fig_continuum} shows the dust continuum emission at 1.3 mm (225 GHz) obtained
with the \com configuration as well as the observations with the \vex~ configuration.
The \com configuration shows a bright source elongated along the north-east south-west direction, similar to 
the torus structure observed by \cite{araya2007} at 7 mm, and a weaker component nearly perpendicular that has the 
same direction of the molecular outflow (see Section~\ref{sect_outflow}). The integrated emission in the whole region above the 3$\sigma$ level is 503 mJy. To obtain the flux associated with the more compact emission, we fitted a 2-D Gaussian using the CASA software package. The measured integrated intensity was 380 mJy with a deconvolved size of 2\farcs2 $\times$ 
1\farcs7 (14,700 $\times$ 11,400 AU at 6.7 kpc).

For the \vex configuration the integrated intensity above 3$\sigma$ is 168 mJy. A 2-D Gaussian fit to the most 
compact emission gives a deconvolved size of 0\farcs8 $\times$ 0\farcs5 (5,400 $\times$ 3,300 AU at 6.7 kpc) and an 
integrated intensity of 125 mJy.  The dust continuum peak  is located  at the same position of the 87 GHz and 46 GHz continuum emission peaks detected by \citet{zhang2007}. We detect no emission toward the secondary 87 GHz peak detected by \cite{zhang2007} at (R.A.,Dec)$_{\mbox{\tiny J2000}}$= (18$^{\mbox{\tiny h}}$59$^{\mbox{\tiny m}}$09.21$^{\mbox{\tiny s}}$,04$\grad$12\arcmin22\farcs6).

\subsection{Spectral Lines}

Figure \ref{fig_sbcompact} shows the observed spectral bandpass (averaged in time and baseline) of the visibility dataset around 220 and 230 GHz of the lower and upper sidebands obtained with  the compact configuration. All the lines identified in the \vex configurations were also detected in the \com configuration (see Table~\ref{tb_lines} for details).

  The spectra reveal a chemically-rich core with 31 lines from 14 species with a broad range of energy levels (11.0 K for $\trece$(2--1) up to 525.6 K for CH$_{3}$CN(12$_{8}$--11$_{8}$)).  The identification of lines is based on frequencies and line strengths listed in the Splatalogue Spectral line database.\footnote{http://www.splatalogue.net/}
 The lines detected include three CO isotopologues ($^{12}$CO, $^{13}$CO, and C$^{18}$O), sulfur-bearing species (SO, 
 O$^{13}$CS, OCS, and $^{13}$CS), and the dense gas tracers CH$_{3}$CN and CH$_{3}$OH  that usually arise from hot cores. 

\begin{figure*}[!htbp]
\begin{center}
\includegraphics[angle=0,scale=1.0]{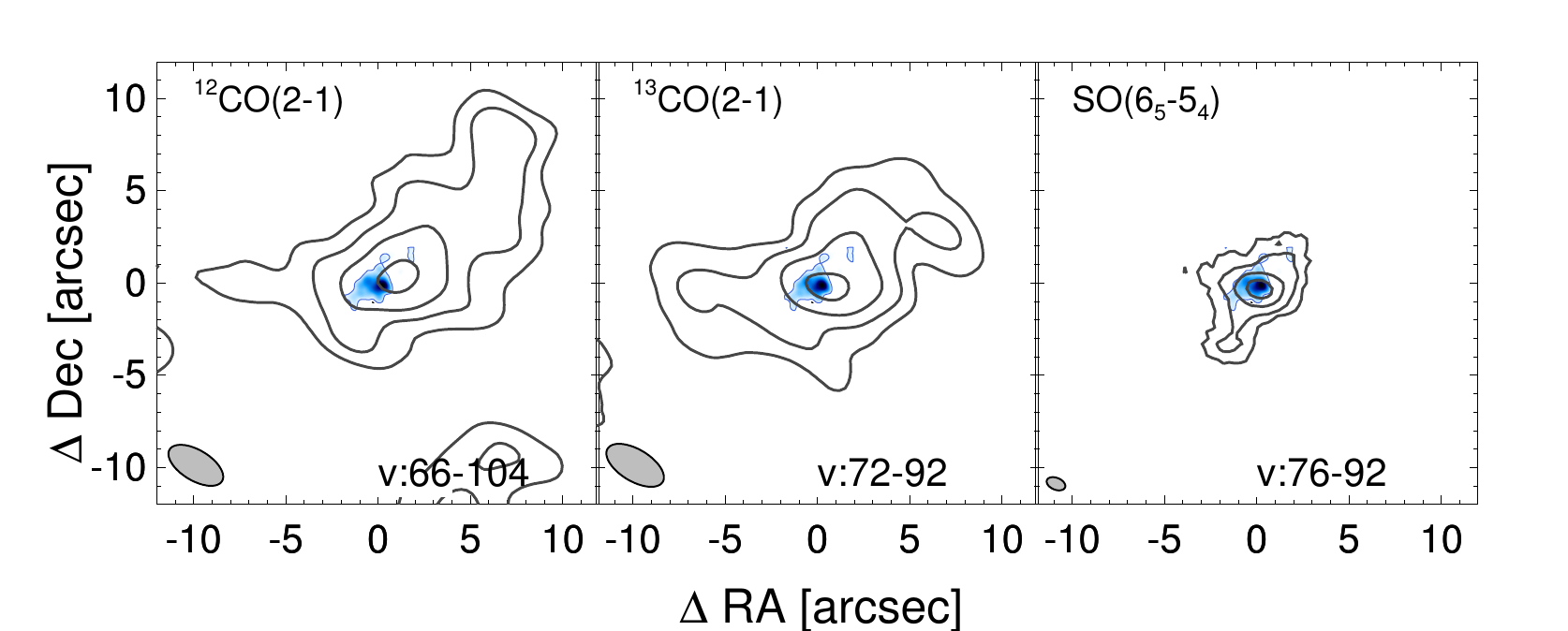}
\caption{ Integrated intensity maps of detected molecular lines with emission at high velocities. The contour levels start at 3$\sigma$ and are in steps of 3$\sigma$ ($\sigma$ values are shown in Table \ref{tb_lines}). The range of integration (in km s$^{-1}$) is indicated at the bottom right corners.  These maps are created using only SMA data.  \label{fig_spectra_wide}}
\end{center}
\end{figure*}

 We produce the integrated intensity maps of all the detected lines. For lines identified in both configurations, we make integrated maps combining the two datasets. 
Most of the species show compact spatial emission coincident with the dust continuum (Figure \ref{fig_spectra_compact_1} and \ref{fig_spectra_compact_2}). 
 However, some of them show elongated emission in the same direction of the continuum bridge (Figure  \ref{fig_spectra_wide}). This direction is the same of the molecular outflow discussed in Section~\ref{sect_outflow}.  
 Table \ref{tb_lines} lists the molecular and observational parameters of  all the detected lines.

\begin{figure*}[!htbp]
\begin{center}
\includegraphics[angle=0,scale=0.7]{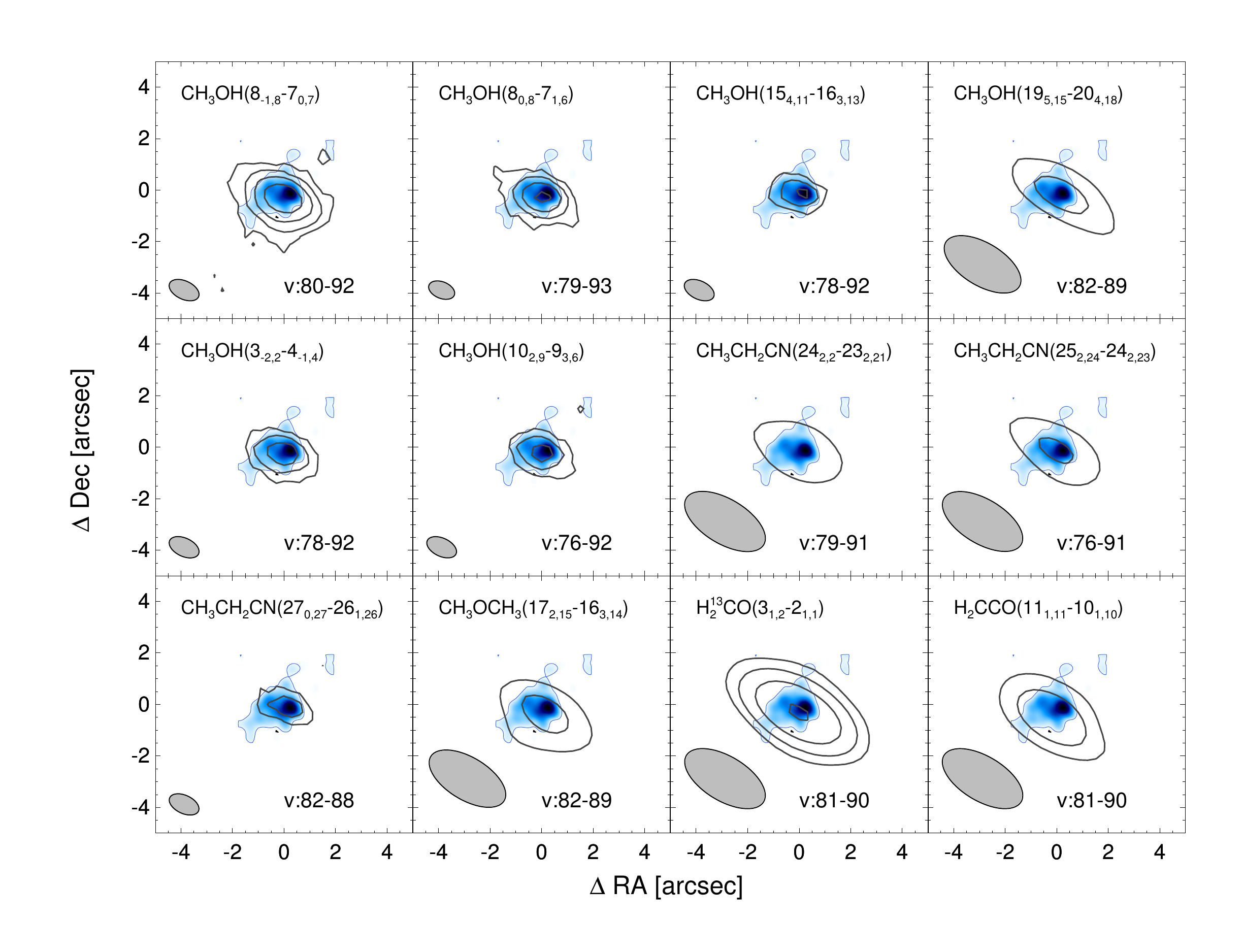}
\caption{ Integrated intensity maps of the detected molecular transitions. The contour levels start at 3$\sigma$ and are in steps of 6$\sigma$ (see Table \ref{tb_lines} for the $\sigma$ values).  The name of the molecular lines are shown at the top of the images.
 The synthesized beams and  range of integration (in km s$^{-1}$)  are shown at the bottom of each panel.  Large beams correspond to images obtained with the $\com$ configuration ($\sim$2\farcs4), while smaller beams correspond to those obtained combining the $\com$+ $\vex$ configurations (0\farcs9).  Blue filled contours correspond to the continuum emission obtained with the very-extended configuration.  \label{fig_spectra_compact_1} 
 }
\end{center}
\end{figure*}

\begin{figure*}[!htbp]
\begin{center}
\includegraphics[angle=0,scale=0.7]{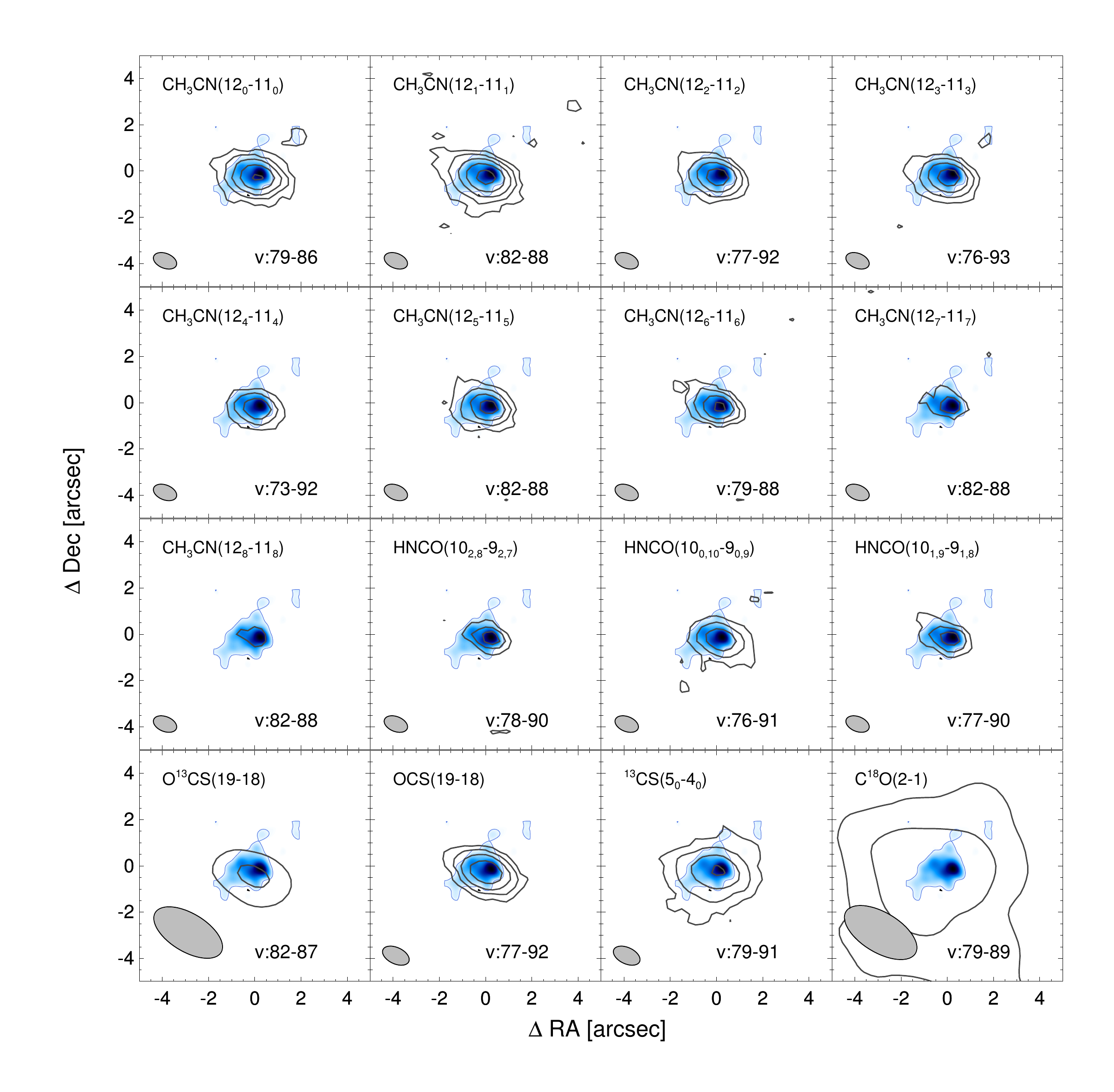}
\caption{ Continuation Figure \ref{fig_spectra_compact_1}.  \label{fig_spectra_compact_2} 
 }
\end{center}
\end{figure*}

%=====================================================================
\subsection{Temperature}\label{sect_temperature}

\begin{figure}[!htbp]
\begin{center}
\includegraphics[angle=0,scale=0.28]{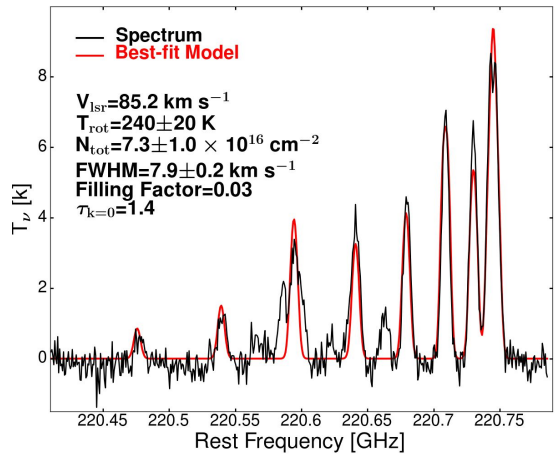}
\caption{The spectra and best-fit model of the CH$_{3}$CN(12$_{k}$-11$_{k}$)  $k$=0-8 lines.  The black line shows the observed CH$_{3}$CN transitions while the red line shows the synthetic spectra obtained from the best-fit model. The resultant parameters from the fit (central velocity, rotational temperature, total column density, full width at half maximum, filling factor, and opacity) are indicated. \label{fig_rotdiag} }
\end{center}
\end{figure}

\begin{figure}[!htbp]
\begin{center}
\includegraphics[angle=0,scale=0.35]{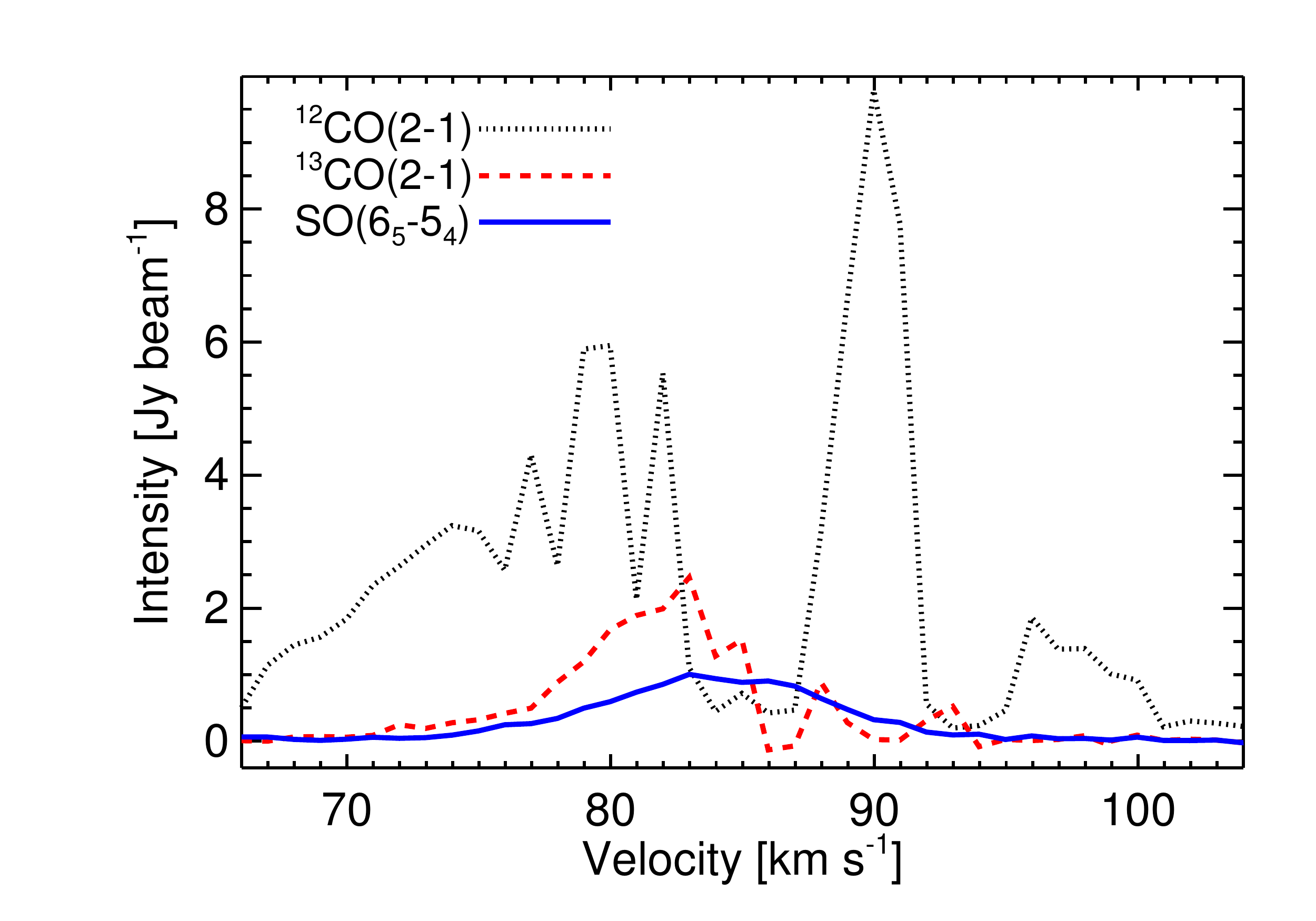}
\caption{ Spectra of the $\dcodu$, $\tcodu$, and $\sosccc$   lines toward the position of the dust peak. The lines have widths up  to 50 km s$^{-1}$. \label{fig_central_spectra}}
\end{center}
\end{figure}

We detect nine components of the $\chcn$ $k$-ladder with $k$=0-8 (Table \ref{tb_lines}).
CH$_{3}$CN is a symmetric top molecule, with a  large line strength and dipole moment (3.91 D). The  CH$_{3}$CN(12$_{k}$--11$_{k}$) level is well-suited for measuring the gas temperature in hot cores \citep[e.g.,][]{pankonin2001, arayita2005, qiu2009}. Since radiative processes across $k$-ladders are forbidden, the population in the different $k$-ladders can be used to estimate the kinetic temperature even if the levels are not fully thermalized.  

We have simultaneously fit all CH$_{3}$CN transitions with the simple model described  by \cite{arayita2005} and \citet{zhang1998}  assuming 
local thermodynamic  equilibrium (LTE) conditions\footnote{The Python script used for fitting the CH$_{3}$CN(12$_{k}$--11$_{k}$) transitions and determining the rotational temperature and column density can be obtained from the following 
link: https://github.com/xinglunju/emanon}. The separation of the different transitions is restricted by their 
rest frequency and we have assumed that all lines have the same velocity dispersion and filling factor. The free parameters in the fitting are 
the rotational temperature, column density, line width, and filling factor. The optical depth was calculated using the best-fit temperature, column density, and line width, following Equation A8 of \cite{arayita2005} (we detected no isotopologue of CH$_{3}$CN).

The CH$_{3}$CN spectra and the best-fit model are shown in Figure~\ref{fig_rotdiag}. The best fit resulted in a 
rotational temperature, $T_{\rm rot}$, of 240 $\pm$ 20 K and a column density, N$_{\rm tot}$, of (7.3 $\pm$ 1.0) $\times$
 10$^{16}$ cm$^{-2}$, values similar to those found in other hot cores \citep{galvan2010, cesaroni2011, qiu2012}.

%=====================================================================
\subsection{Mass of the Core}

\citet{zhang2007} derived a spectral index of $\beta$=1.3 by fitting a greybody to the SED of \source, indicating that the entire millimeter continuum from the source is attributable to thermal emission from dust. 
We adopt the rotational temperature T=45 K derived by \citet{zhang2007} using NH$_{3}$ line observations as the dust temperature in the region since this is more representative of the extension of the dust emission in the compact configuration.  
For optically thin and isothermal emission, the mass of the core can be obtained from the expression  
\begin{equation}
M_{gas}=\frac{S_{1.3mm}D^{2}}{R_{dg}k_{1.3mm}B_{1.3mm}(T_{d})},
\end{equation}
where $S_{1.3mm}$ is the integrated flux density at 1.3 mm,
 $D$ is the distance to the source,
$R_{dg}$ is the dust-to-gas mass ratio with a value of 0.01,
$B_{1.3mm}(T_{d}$) is the Planck function at the dust temperature $T_{d}$, and
$k_{1.3mm}$ is the dust absorption coefficient  assumed to be equal to 
0.9 cm$^{2}$ g$^{-1}$ at 230 GHz \citep{ossenkopf1994}.

Assuming T=45 K and using the integrated flux from the 2-D Gaussian fitting,  we obtain a gas mass of 150  and 50 $\msun$ for the  $\com$ and the \vex configuration, respectively. Using the temperature of 240 K derived from the $k$-ladders of the CH$_{3}$CN(12-11) line, the measured mass for the \vex configuration is 8 $\msun$.  
%Assuming a spherical hot core, the volume and surface densities are 2.6 $\times$ 10$^7$ cm$^{-3}$ and 5.2 g~cm$^{-2}$. 
This mass of 8 $\msun$ corresponds to the mass of the central, dense compact ($\sim$4,000 AU) hot core.

\begin{figure*}[!htbp]
\begin{center}
\includegraphics[angle=0,scale=0.99]{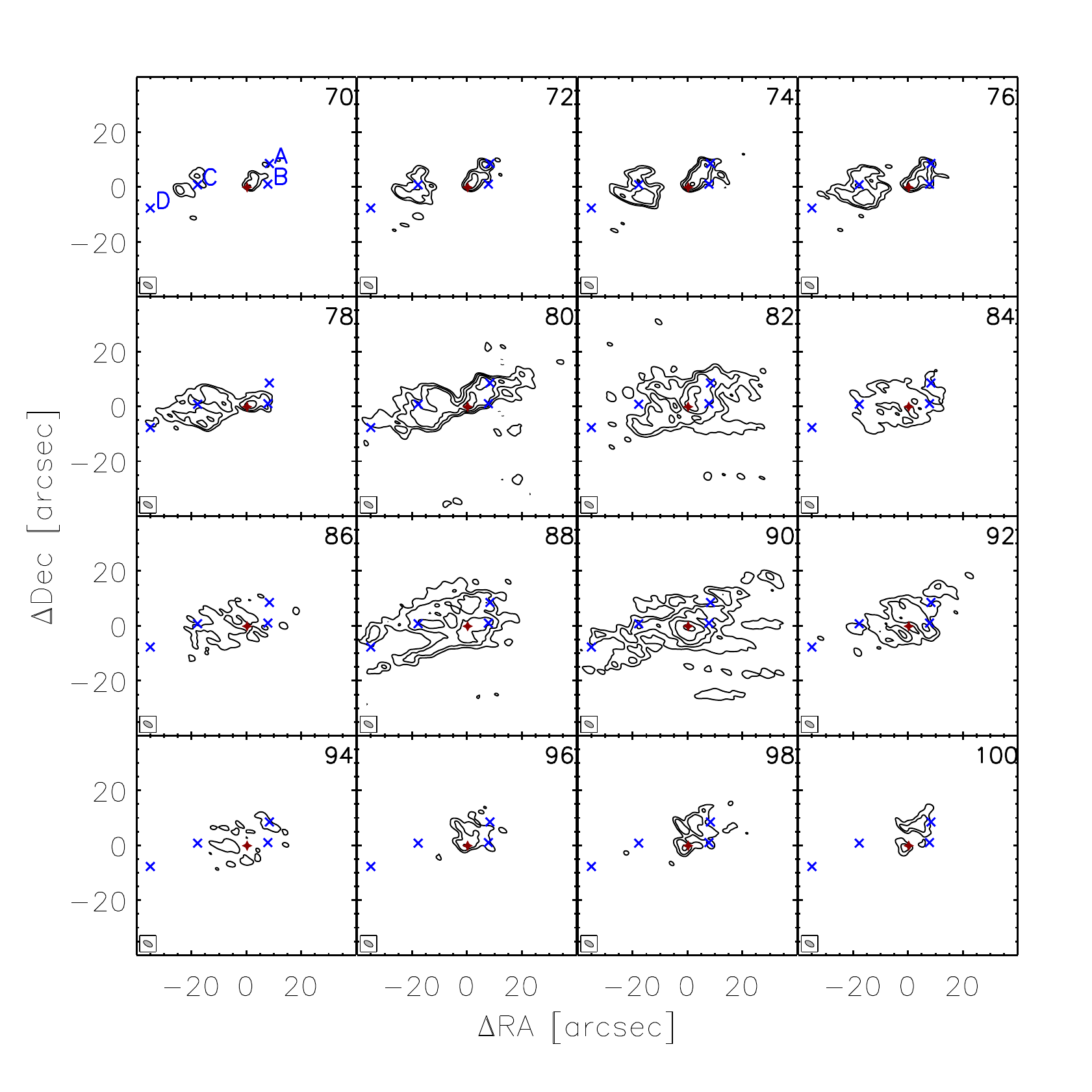}
\caption{ Channel map of the SMA+IRAM observations of the $\dcodu$ line. The velocities are indicated at the top right corner in km s$^{-1}$.
The beam size is shown at the bottom left corner.   The contours start at 9$\sigma$  and are in steps of 9$\sigma$ (1$\sigma$=60 mJy beam$^{-1}$).  The NH$_{3}$(3,3) features detected by \citet{zhang2007} are indicated with the blue crosses while the dust peak is indicated with the red star. \vspace{0.4cm} 
\label{fig_channel_co21} }
\end{center}
\end{figure*}

%=====================================================================

\subsection{Molecular Outflow}\label{sect_outflow}

\begin{figure*}[!htbp]
\begin{center}
\includegraphics[angle=0,scale=0.6]{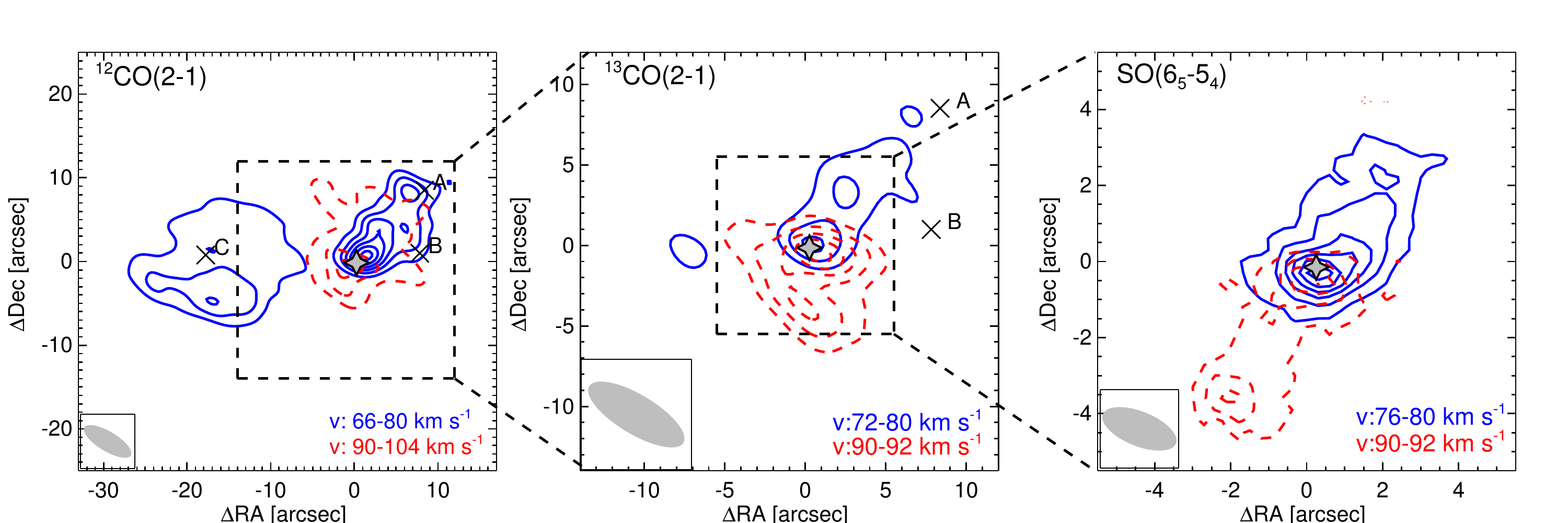}
\caption{  From left to right, blueshifted and redshifted emission obtained for the $\dcodu$,  $\tcodu$, and $\sosccc$ lines, respectively.  The dashed boxes indicate the region for more compact emission shown at the right side.
The contour levels start at 9, 6 and 3$\sigma$ and are in steps of 9, 6, and 3$\sigma$ for $\dcodu$, $\tcodu$, and $\sosccc$, respectively.
 The range of velocity integration is  indicated at the bottom right in each box. The beam sizes are shown at the bottom left corners. The NH$_{3}$(3,3) features detected by \citet{zhang2007} are indicated with the crosses while the dust peak is indicated with the gray star. \label{fig_outflows}\\ \\ \\}
\end{center}
\end{figure*}

\begin{figure*}[!htbp]
\begin{center}
\includegraphics[angle=0,scale=0.9]{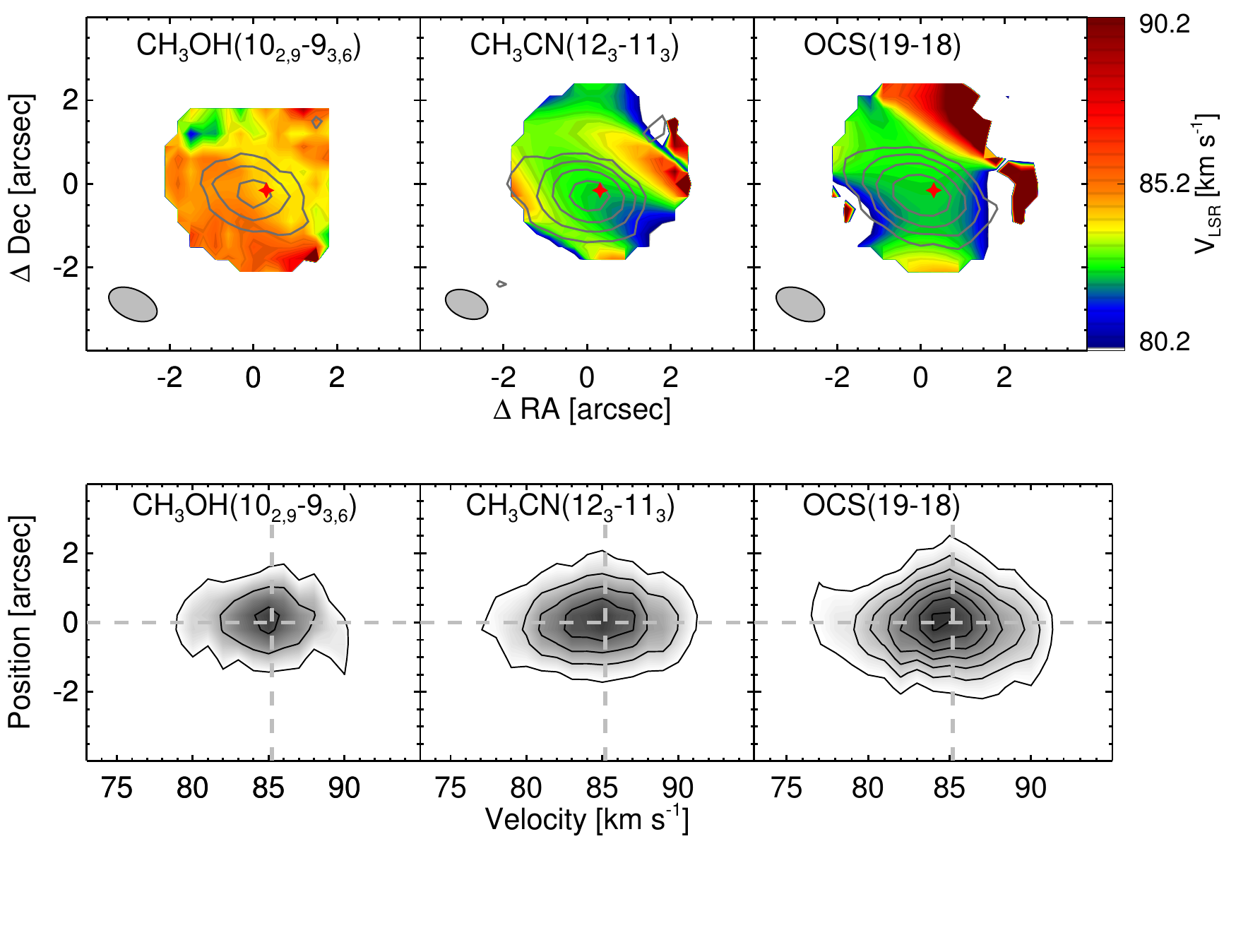}
\caption{ {\it Top:} Moment one maps  (color scale) for some of the lines with the highest energy levels (Table \ref{tb_lines}) overplotted with the moment zero maps (gray lines). {\it Bottom:} The position velocity maps for the respective lines. We find no clear trend for rotational structures. \label{fig_pv_mom_rot}}
\end{center}
\end{figure*}

The $\dcodu$, $\tcodu$,  and $\sosccc$ spectra show emission in an  extended velocity range
 (Figure \ref{fig_central_spectra}).
 The CO isotopologue spectra  show dips around the systemic velocity $\vlsr$= 85.2 $\kms$. The dips are  likely caused by both self-absorption  and  missing flux of the extended cloud envelope in the interferometer data. The interferometric spatial filtering affects the $\dcodu$ line strongest since it is the most abundant isotopologue and traces the most extended regions that cannot be covered by the shortest baselines of the interferometer.
 We recover the missing flux of $\dcodu$ by combining in the visibility domain the single-dish data from the IRAM 30 m telescope and the interferometric data  \citep{zhang2000}.   
 
 Figure \ref{fig_channel_co21}  shows the channel maps produced using  the SMA+IRAM observations of $\dcodu$. 
The channel maps show emission extending in north-west south-east direction in a velocity range of $\Delta$v = 30 km s$^{-1}$, tracing a molecular outflow in the same direction as the bipolar molecular outflow delineated in SiO 
emission by \citet{zhang2007}. Figure~\ref{fig_outflows} shows the integrated intensity maps of the wing emission for the $\dcodu$, $\tcodu$, and $\sosccc$ lines.  We excluded in the integration the velocity range associated with the ambient cloud (81--89 $\kms$), as defined by molecular emission that seems unaffected by the outflow. 

 The outflow maps reveal emission centered at the position of the dust core.  
  Using the $\dcodu$ line from the combined interferometric and single-dish data set,  we compute mass, momentum, and energy of the outflow following the procedure described in \citet{cabrit1986}. We assume optically-thin emission and an excitation temperature of 70 K.  With a total mass of 16.8 
  $\msun$, the molecular outflow is consistent with other molecular outflows found in high-mass star-forming regions 
  \cite[e.g.,][]{beuther2002b, zhang2001, zhang2005}.
  It has a mass $\sim$60\% lower than the 27 
  $\msun$ derived by \citet{zhang2007} using SiO. This discrepancy can be produced by the more uncertain conversion factor from SiO to H$_2$ column than from CO to H$_2$, and by the assumption of optically thin CO emission. Indeed, CO wing emission with an optical depth of 1 would increase the mass to the same value measured using SiO.  The parameters of the outflow are shown in Table \ref{tb_outflow}.

\citet{zhang2007} identified four compact and narrow line NH$_{3}$(3,3) emission features that likely arise from a weak population inversion (labeled as ``A", ``B", ``C", and ``D" in Figure~\ref{fig_channel_co21} and \ref{fig_outflows}). 
They suggest that these NH$_{3}$(3,3) features are associated to the outflow activity in the region and that ``C" and ``D" could be part of an outflow driven by an undetected core 
located between them. We detect no dust continuum core (nor hot core line emission) between ``C" and ``D" and no hint of a molecular outflow 
other than the one driven by the bright dust continuum core. Given the position of these features in the channel 
map of CO, we conclude that they are all associated with the outflow emanating from the massive dust core.

\begin{deluxetable}{lccc}
\tabletypesize{\normalsize}
\tablecaption{Parameters of the outflow derived from the $\dcodu$ emission.\label{tb_outflow}}
\tablewidth{0pt}
\tablehead{
\colhead{}&\colhead{Mass} &\colhead{Momentum} &\colhead{Energy} \\
\colhead{Lobe}&\colhead{($\msun$)} &\colhead{($\msun ~ \kms$)} &\colhead{($\msun$ km$^{2}$ s$^{-2}$)}
}
\startdata 
Blue  & 11.4  & 105 & 600 \\
Red & 5.4     & 45 & 240\\
Total & 16.8  & 150 & 840 \\
\enddata
\end{deluxetable}

%=====================================================================
\subsection{Velocity gradients}\label{sect_rotation}

We investigated the  velocity gradients for all the detected lines in \source\ and find no clear trend for rotational structures at 
4,000 AU scales. As an example, we present in Figure~\ref{fig_pv_mom_rot} the moment one and the position-velocity maps for three spectral lines 
(CH$_{3}$OH(10$_{2,9}$-9$_{3,6}$)A, CH$_{3}$CN(12$_{3}$-11$_{3}$), and OCS(19-18)). Position-velocity maps were made in a direction roughly perpendicular to the direction of the molecular outflow. The absence of velocity gradients 
does not rule out the existence of an accretion disk (sizes $<$1,000 AU) feeding the central high-mass star(s), but discards the 
presence of a toroidal-rotating structure at several thousand AU scales.

%=====================================================================

\section{Discussion and Conclusion}\label{sect_discuss}

We detected a variety of spectral lines as well as 1.3 mm continuum emission which allowed us to measure the mass and temperature of the compact core embedded in \source. Molecular emission at high-velocities (e.g., CO and SO) reveal a bipolar molecular outflow for which we determine its properties. This molecular outflow is roughly perpendicular to the torus-like structure detected by \cite{araya2007} at 7 mm continuum (also seen at the low angular resolution with SMA). However, molecular tracers often used to study core rotation (e.g., CH$_3$CN) show no clear signs of rotating structures.

Theories of massive star formation predict that outflows must be powered and fed by accretion disks. Evidence of rotation and infall that hint at the presence of disks has been detected in several systems \citep[eg.,][]{cesaroni1997, zhang1998, cesaroni2005b, beuther2007, zhang2009, fallscheer2011, qiu2012, cesaroni2011, beltran2011, guzman2014,Johnston15,Chen16,Ilee16,Beuther17}. 
However, the velocity structures shown in Figure \ref{fig_pv_mom_rot}  indicate that there is no dominant angular momentum
axis in the hot core at a spatial scale of 4000 AU. The size scale of the accretion disk responsible for driving the CO outflow must be much smaller. Small disks are expected in dense cluster forming regions as cores harboring massive stars tend to fragment and form multiple stars. The dynamical interactions limit the size of circumstellar disks. In addition, the complex velocity structures seen in the CH$_3$OH, CH$_3$CN and OCS emission may also be partly due to the influence of the molecular outflow, which affects the kinematics and chemistry of the gas.

%However, it is intriguing why there are no detectable signs of rotation in \source. We attribute the lack of rotation at 4,000 AU scales to probably the combination of both (i) the multiplicity at smaller scales and (ii) the contamination of the molecular outflow in the observed molecular transitions. 

 High-mass stars are overwhelmingly observed in binary or multiple systems. Indeed, in a high-resolution spectroscopic survey
 including nearly 800 O- and B-type stars, 82\% of high-mass stars were found in close binary systems \citep{Chini12}.
  In addition, the majority of high-mass stars are paired with high-mass companions \citep{Chini12,Kobulnicky07}. 
  We cannot rule out the formation of a binary or multiple system embedded in the \source\ hot core. By comparing the 
  structure of the continuum emission observed with the compact and very-extended configurations, we can see that 
  some substructures are detected, but not resolved. It is possible that rotational motions are present on smaller scales unaccessible with our SMA observations. 
  Indeed, in the nearby high-mass star-forming region NGC 6334I, at least 4 high-mass stars are in the 
  process of formation in less that 2,500 AU \citep{Brogan16}. 
  Higher spatial resolution will be needed to search for possible multiplicity in \source. 

\cite{Leurini11} and recently \cite{Beuther17} suggest that in IRAS 17233-3606 the CH$_3$CN emission, typically 
used to reveal rotating structures (toroid-disk-like), can be affected by molecular outflows. This is particularly interesting 
in the study of  \cite{Beuther17}  at 0\farcs06 resolution (130 AU scales), in which the P-V diagrams show Keplerian motions 
and contamination that is attributed to both infall and molecular outflows even in high-excitation CH$_3$CN(37$_k$-36$_k$) transitions 
($E_u$$\sim$1,000 K). Therefore, high angular resolution observations in \source\ are not only necessary to resolve possible 
further fragmentation, but also to clarify if CH$_3$CN and the other observed hot core tracers have contamination 
from processes other than rotation that are unresolved at the SMA angular resolution, precluding the detection of disks. 
Interestingly, the recent observations of CH$_3$CN(5-4) in \source\ at 1\arcsec\ resolution by \cite{Hofner17} show a 
velocity gradient in the direction of the jet/outflow. Their observations support the idea that CH$_3$CN emission can 
be contaminated by outflowing gas. This velocity gradient is likely not evident in our CH$_3$CN(12-11) lines because the 
higher transitions will be excited at higher temperatures, closer to the high-mass star and confusion from both 
physical processes, rotation and outflowing gas, can be more important than at lower energy transitions.  This is supported by the lower filling factor of 0.03 derived in this work for CH$_3$CN(12-11) with respect to 0.17 derived 
by  \cite{Hofner17} for  CH$_3$CN(5-4). 

In summary, we have examined the continuum and spectral line emission and draw the following conclusions:

1. Using SMA in the compact and very extended configurations, we have observed 1.3 mm continuum and molecular line emission  toward $\source$ to reveal a chemically-rich, compact core with a  mass of 8 $\msun$ (50 $\msun$ at 45 K) and central temperature of 240 K that drives a bipolar molecular outflow with a total mass of 16.8 M$_{\odot}$. 

2. The molecular outflow in CO and SO is consistent with the previously detected outflow in SiO. The molecular outflow 
is roughly perpendicular to the torus structure detected in continuum emission with the compact configuration \citep[also 
detected by][]{araya2007}. However, this torus shows substructures at higher angular resolution and present 
no velocity gradients that would suggest a toroidal-disk-like rotating structure. 

3. We suggest that the lack of rotational structures is due to multiplicity of high-mass stars embedded in the hot 
core and possibly the influence of the outflow (and maybe infall) in the hot core tracers used to search for 
velocity gradients. Higher angular resolution, likely with ALMA, can confirm or refute both scenarios.

\acknowledgements
We  thank the anonymous referee for helpful comments that improved the paper.
Data analysis was in part carried out on the open use data analysis computer system 
at the Astronomy Data Center, ADC, of the National Astronomical Observatory of Japan.

\vspace{5mm}
\facilities{SMA, IRAM}

\software{MIRIAD \citep{sault1995}, Python\footnote{https://www.python.org}, emanon\footnote{https://github.com/xinglunju/emanon}, CASA \citep{mcmullin2007}, CLASS\footnote{http://www.iram.fr/IRAMFR/GILDAS}
          }

%=====================================================================


\begin{thebibliography}{}

\bibitem[Al-Marzouk et al.(2012)]{almarzouk2012} Al-Marzouk, A.~A., Araya, E.~D., Hofner, P., et al.\ 2012, \apj, 750, 170 

\bibitem[Araya et al.(2005a)]{arayita2005} Araya, E., Hofner, P., Kurtz, S., Bronfman, L., \& DeDeo, S.\ 2005, \apjs, 157, 279 

\bibitem[Araya et al.(2005b)]{araya2005} Araya, E., Hofner, P., Kurtz, S., et al.\ 2005, \apj, 618, 339 

\bibitem[Araya et al.(2007)]{araya2007} Araya, E., Hofner, P., Sewi{\l}o, M., et al.\ 2007, \apj, 669, 1050 

\bibitem[Beltr{\'a}n et al.(2011)]{beltran2011} Beltr{\'a}n, M.~T., Cesaroni, R., Zhang, Q., et al.\ 2011, \aap, 532, A91 

\bibitem[Beltr{\'a}n \& de Wit(2016)]{Beltran16} Beltr{\'a}n, M.~T., \& de Wit, W.~J.\ 2016, \aapr, 24, 6

\bibitem[Beuther et al.(2002a)]{beuther2002a} Beuther, H., Schilke, P., Menten, K.~M., et al.\ 2002a, \apj, 566, 945 

\bibitem[Beuther et al.(2002b)]{beuther2002b} Beuther, H., Schilke, P., Sridharan, T.~K., et al.\ 2002b, \aap, 383, 892 

\bibitem[Beuther et al.(2002c)]{beuther2002c} Beuther, H., Walsh, A., Schilke, P., et al.\ 2002c, \aap, 390, 289 

\bibitem[Beuther et al.(2005)]{beuther2005} Beuther, H., Zhang, Q., Greenhill, L.~J., et al.\ 2005, \apj, 632, 355 

\bibitem[Beuther et al.(2006)]{beuther2006} Beuther, H., Zhang, Q., Reid, M.~J., et al.\ 2006, \apj, 636, 323 

\bibitem[Beuther et  al.(2007)]{beuther2007} Beuther, H., Zhang, Q., Bergin, E.~A., et al.\ 2007, \aap, 468, 1045 

\bibitem[Beuther et al.(2017)]{Beuther17} Beuther, H., Walsh, A.~J., Johnston, K.~G., et al.\ 2017, arXiv:1703.07235

\bibitem[Brogan et al.(2016)]{Brogan16} Brogan, C.~L., Hunter, T.~R., Cyganowski, C.~J., et al.\ 2016, \apj, 832, 187

\bibitem[Bronfman et al.(1996)]{bronfman1996} Bronfman, L., Nyman, L.-A., \& May, J.\ 1996, \aaps, 115, 81 

\bibitem[Cabrit \& Bertout(1986)]{cabrit1986} Cabrit, S., \& Bertout, C.\ 1986, \apj, 307, 313 

\bibitem[Carral et al.(1999)]{carral1999} Carral, P., Kurtz, S., Rodriguez, L.~F., et al.\ 1999, RMxAA, 35, 97 

\bibitem[Cesaroni et al.(1997)]{cesaroni1997} Cesaroni, R., Felli, M., Testi, L., Walmsley, C.~M., \& Olmi, L.\ 1997, \aap, 325, 725 

\bibitem[Cesaroni(2005a)]{cesaroni2005a} Cesaroni, R.\ 2005, Massive Star Birth: A Crossroads of Astrophysics, 227, 59 

\bibitem[Cesaroni et al.(2005b)]{cesaroni2005b} Cesaroni, R., Neri, R., Olmi, L., et al.\ 2005, \aap, 434, 1039 

\bibitem[Cesaroni et al.(2011)]{cesaroni2011} Cesaroni, R., Beltr{\'a}n, M.~T., Zhang, Q., Beuther, H., \& Fallscheer, C.\ 2011, \aap, 533, A73 

\bibitem[Chen et al.(2016)]{Chen16} Chen, H.-R.~V., Keto, E., Zhang, Q., et al.\ 2016, \apj, 823, 125

\bibitem[Chini et al.(2012)]{Chini12} Chini, R., Hoffmeister, V.~H., Nasseri, A., Stahl, O., \& Zinnecker, H.\ 2012, \mnras, 424, 1925 

\bibitem[Cyganowski et al.(2012)]{cyganowski2012} Cyganowski, C.~J.,  Brogan, C.~L., Hunter, T.~R., et al.\ 2012, IAU Symposium, 287, 127 

\bibitem[Fallscheer et al.(2011)]{fallscheer2011} Fallscheer, C.,  Beuther, H., Sauter, J., Wolf, S., \& Zhang, Q.\ 2011, \apj, 729, 66 

\bibitem[Galv{\'a}n-Madrid et al.(2010)]{galvan2010}  Galv{\'a}n-Madrid, R., Zhang, Q., Keto, E., et al.\ 2010, \apj, 725, 17 

\bibitem[Garay \& Lizano(1999)]{Garay1999} Garay, G., \& Lizano, S.\ 1999, \pasp, 111, 1049

\bibitem[Garrod \& Herbst(2006)]{garrod2006} Garrod, R.~T., \& Herbst, E.\ 2006, \aap, 457, 927 

\bibitem[Guzm{\'a}n et al.(2014)]{guzman2014} Guzm{\'a}n, A.~E., Garay, G., Rodr{\'{\i}}guez, L.~F., et al.\ 2014, \apj, 796, 117 

\bibitem[Guzm{\'a}n et al.(2016)]{Guzman16} Guzm{\'a}n, A.~E., Garay, G., Rodr{\'{\i}}guez, L.~F., et al.\ 2016, \apj, 826, 208 

\bibitem[Ho et al.(2004)]{ho2004} Ho, P.~T.~P., Moran, J.~M., \& Lo, K.~Y.\ 2004, \apjl, 616, L1 

\bibitem[Hofner et al.(2017)]{Hofner17} Hofner, P., Cesaroni, R., Kurtz, S., et al.\ 2017, arXiv:1705.07203

\bibitem[Ilee et al.(2016)]{Ilee16} Ilee, J.~D., Cyganowski, C.~J., Nazari, P., et al.\ 2016, \mnras, 462, 4386

\bibitem[Johnston et al.(2015)]{Johnston15} Johnston, K.~G., Robitaille, T.~P., Beuther, H., et al.\ 2015, \apjl, 813, L19 

\bibitem[Keto(2002)]{keto2002} Keto, E.\ 2002, \apj, 580, 980 

\bibitem[Keto(2003)]{keto2003} Keto, E.\ 2003, \apj, 599, 1196 

\bibitem[Kobulnicky \& Fryer(2007)]{Kobulnicky07} Kobulnicky, H.~A., \& Fryer, C.~L.\ 2007, \apj, 670, 747 

\bibitem[Kurtz et al.(2000)]{kurtz2000} Kurtz, S., Cesaroni, R., Churchwell, E., Hofner, P., \& Walmsley, C.~M.\ 2000, Protostars and Planets IV, 299 

\bibitem[Leurini et al.(2011)]{Leurini11} Leurini, S., Codella, C., Zapata, L., et al.\ 2011, \aap, 530, A12 

\bibitem[McMullin et al.(2007)]{mcmullin2007} McMullin, J.~P., Waters, B., Schiebel, D., Young, W., \& Golap, K.\ 2007, Astronomical Data Analysis Software and Systems XVI, 376, 127 


\bibitem[Miralles et al.(1994)]{miralles1994} Miralles, M.~P., Rodriguez, L.~F., \& Scalise, E.\ 1994, \apjs, 92, 173 

\bibitem[Molinari et al.(1996)]{molinari1996} Molinari, S., Brand, J., Cesaroni, R., \& Palla, F.\ 1996, \aap, 308, 573 

\bibitem[Ossenkopf \& Henning(1994)]{ossenkopf1994} Ossenkopf, V., \& Henning, T.\ 1994, \aap, 291, 943 

\bibitem[Pankonin et al.(2001)]{pankonin2001} Pankonin, V., Churchwell, E., Watson, C., \& Bieging, J.~H.\ 2001, \apj, 558, 194 

\bibitem[Qiu et al.(2008)]{qiu2008} Qiu, K., Zhang, Q., Megeath, S.~T., et al.\ 2008, \apj, 685, 1005 

\bibitem[Qiu \& Zhang(2009)]{qiu2009} Qiu, K., \& Zhang, Q.\ 2009, \apjl, 702, L66 

\bibitem[Qiu et al.(2012)]{qiu2012} Qiu, K., Zhang, Q.,  Beuther, H., \& Fallscheer, C.\ 2012, \apj, 756, 170 

\bibitem[Sanchez-Monge et al.(2017)]{sanchez17} Sanchez-Monge, A., Schilke, P., Schmiedeke, A., et al.\ 2017, arXiv:1704.01805 

\bibitem[Sanhueza et al.(2012)]{sanhueza2012} Sanhueza, P., Jackson, J.~M., Foster, J.~B., et al.\ 2012, \apj, 756, 60

\bibitem[Sault et al.(1995)]{sault1995} Sault, R.~J., Teuben,  P.~J., \& Wright, M.~C.~H.\ 1995, Astronomical Data Analysis Software and Systems IV, 77, 433 

\bibitem[Slysh et al.(1999)]{slysh1999} Slysh, V.~I., Val'tts, I.~E., Kalenskii, S.~V., et al.\ 1999, \aaps, 134, 115 

\bibitem[Sridharan et al.(2002)]{sridharan2002} Sridharan, T.~K., Beuther, H., Schilke, P., Menten, K.~M., \& Wyrowski, F.\ 2002, \apj, 566, 931 

\bibitem[van Dishoeck \& Blake(1998)]{vandishoeck1998} van Dishoeck, E.~F., \& Blake, G.~A.\ 1998, \araa, 36, 317 

\bibitem[Walmsley et al.(1995)]{walmsley1995} Walmsley, C.~M., Cesaroni, R., Olmi, L., Churchwell, E., \& Hofner, P.\ 1995, \apss, 224, 173 

\bibitem[Wang et al.(2006)]{wang2006} Wang, Y., Zhang, Q.,  Rathborne, J.~M., Jackson, J., \& Wu, Y.\ 2006, \apjl, 651, L125 

\bibitem[Wang et al.(2014)]{wang2014} Wang, K., Zhang, Q., Testi, L., et al.\ 2014, \mnras, 439, 3275 

\bibitem[Williams et al.(2004)]{williams2004} Williams, S.~J., Fuller, G.~A., \& Sridharan, T.~K.\ 2004, \aap, 417, 115 

\bibitem[Zhang \& Ho(1995)]{zhang1995} Zhang, Q., \& Ho, P.~T.~P.\ 1995, \apjl, 450, L63 

\bibitem[Zhang et al.(1998)]{zhang1998} Zhang, Q., Ho, P.~T.~P., \& Ohashi, N.\ 1998, \apj, 494, 636 

\bibitem[Zhang et al.(1999)]{zhang1999} Zhang, Q., Hunter, T.~R.,  Sridharan, T.~K., \& Cesaroni, R.\ 1999, \apjl, 527, L117

\bibitem[Zhang et al.(2000)]{zhang2000} Zhang, Q., Ho, P.~T.~P., \& Wright, M.~C.~H.\ 2000, \aj, 119, 1345 

\bibitem[Zhang et al.(2001)]{zhang2001} Zhang, Q., Hunter, T.~R., Brand, J., et al.\ 2001, \apjl, 552, L167 

\bibitem[Zhang et al.(2005)]{zhang2005} Zhang, Q., Hunter, T.~R.,  Brand, J., et al.\ 2005, \apj, 625, 864 

\bibitem[Zhang et al.(2007)]{zhang2007} Zhang, Q., Sridharan, T.~K., Hunter, T.~R., et al.\ 2007, \aap, 470, 269

\bibitem[Zhang et al.(2009)]{zhang2009} Zhang, Q., Wang, Y., Pillai, T., \& Rathborne, J.\ 2009, \apj, 696, 268 

\end{thebibliography}
\end{document}